\documentclass[conference]{IEEEtran}
\IEEEoverridecommandlockouts
\usepackage{cite}
\usepackage{amsmath,amssymb,amsfonts}
\usepackage{algorithmic}
\usepackage{graphicx}
\usepackage{textcomp}
\usepackage{xcolor}
\usepackage{pifont}
\usepackage{hyperref}
\usepackage{cleveref}
\usepackage{flushend}

\def\BibTeX{{\rm B\kern-.05em{\sc i\kern-.025em b}\kern-.08em
    T\kern-.1667em\lower.7ex\hbox{E}\kern-.125emX}}
\begin{document}

\title{An Adaptive Neuro-fuzzy Strategy in Closed-loop Control of Anesthesia \\
}

\author{Mohammad Javad Khodaei$^{1}$, Mohammad Hadi Balaghi Inaloo$^{2}$, Amin Mehrvarz$^{1}$, and Nader Jalili$^{1,3}$ \\
$^{1}$Department of Mechanical Industrial Engineering, Northeastern University, Boston, MA, USA \\
$^{2}$Department of Mechanical Engineering, 
Eindhoven University of Technology, Eindhoven, Netherlands \\
$^{3}$Professor and Head, Department of Mechanical Engineering, University of Alabama, Tuscaloosa, AL, USA
}

\maketitle

\begin{abstract}
This paper proposes an adaptive neuro-fuzzy framework to improve drug infusion rate in closed-loop control of anesthesia. The proposed controller provides a sub-optimal propofol administration rate as input to reach the desired bispectral index, which is the output of the system, in both induction and maintenance phases. In this controller, a critic agent assesses the plant output and produces a reinforcement signal to adapt the controller parameters and minimize the propofol administration rate. The controller is applied to a conventional pharmacokinetic-pharmacodynamics model of anesthesia to evaluate its applicability in closed loop-control of anesthesia. To simulate the designed controller, physiological parameters of 12 patients are used in the mathematical model. The simulation results show that the proposed controller can overcome current challenges in the closed-loop control of anesthesia like inter patient variability, model uncertainties and surgical disturbances without overdose or underdose in a time range of 2 to 4 minutes. Analytical comparison of results shows the strength of the controller in closed-loop control of anesthesia.  
\end{abstract}


\section{Introduction} \label{Introduction}
Biomedical systems are recently the field of interest of both engineers and clinicians since they are needed to be mathematically modeled and controlled. These complex and nonlinear models and controllers improve biologists and clinicians decision-making ability since they predict the physiological system's behavior and make them more sensible. Drug delivery can be taken to account as the main reason for these efforts. A standard and desirable drug delivery system is needed to be automated to prevent additional drug infusion and their side effects as well as improving their sufficient in the desired time. There are several studies in different biologial systems to achive this goal, such as targeted drug delivery in lung for asthmatic subjects\textcolor{blue}{\cite{das2018targeting, zeman2010targeting, poorbahrami2019patient}} or in Kidney to  improve the kidney targeted delivery of angiotensin-converting enzyme\textcolor{blue}{\cite{geng2012peptide}}. Nowadays, several closed-loop controllers are developing to use in different drug delivery systems to improve treatments periods and ways such as glucose regulation of diabetic patients, blood pressure regulation and treatment of neurological disorders.

\begin{figure}
\centering
\includegraphics[width=80mm]{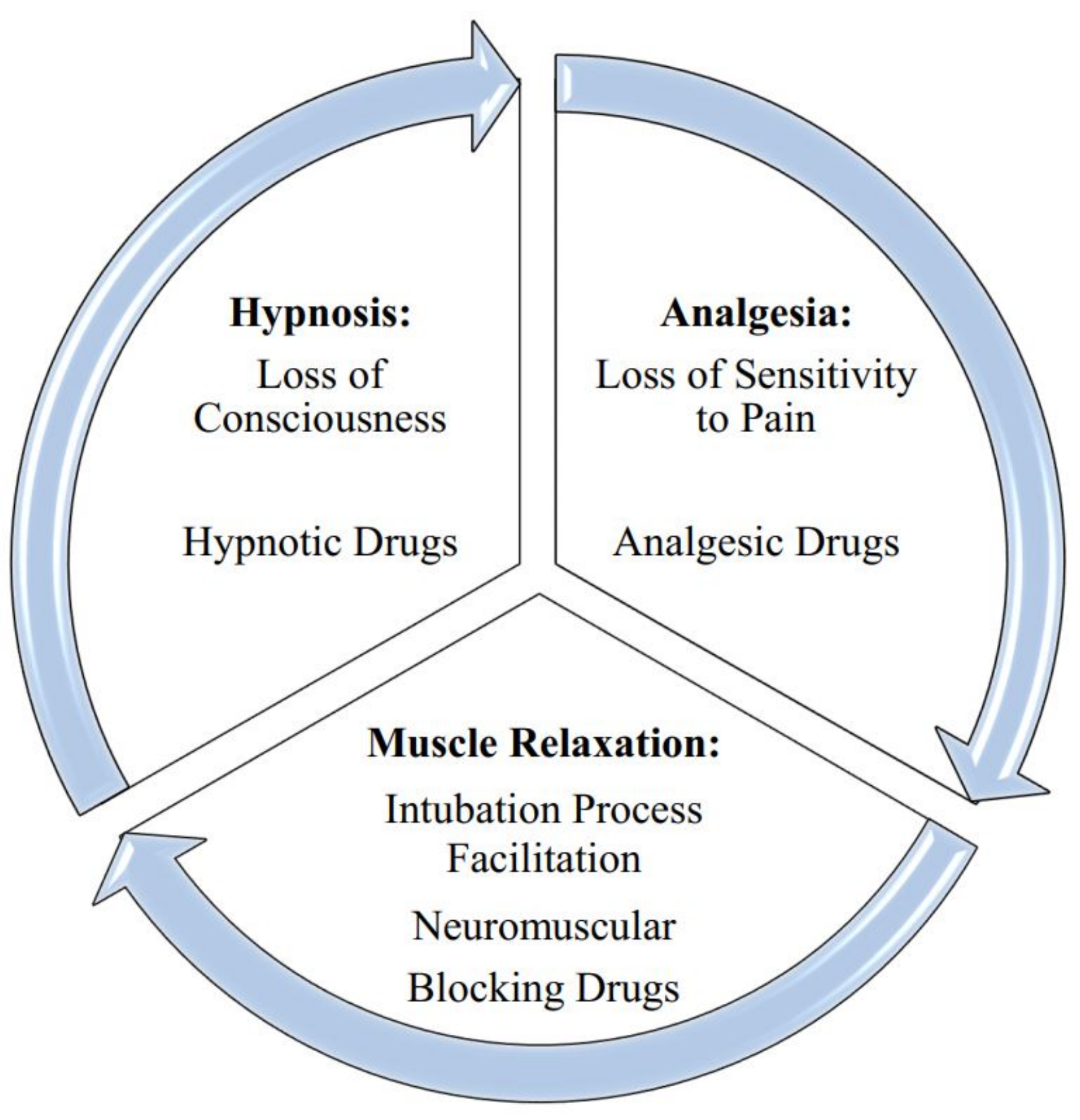}
\caption{The functional components of general anesthesia.}
\label{fig1}
\end{figure}

Anesthesia is one of these biomedical systems that need to be fully understood due to its necessity in most surgical operations. General anesthesia can be obtained by both infusion drugs (intravenous anesthetics) and vapor drugs (inhalational anesthetics). These drugs are in different types and consequently have different rules and effects in anesthesia. As seen in \textcolor{blue}{Fig. \ref{fig1}}, there are three different targets in general anesthesia. The first target is analgesia which can be obtained by analgesic drugs (e.g., remifentanil) administration and indicates that the patients should be at a certain level of consciousness during surgery. It makes the patients unawareness about the stages of surgery and they cannot see the processes, which may have negative mental effects on them. The second necessary part of anesthesia is hypnosis. Hypnosis defines that the patients should not be able to sense any pain in the surgery. This is one of the important parts of anesthesia since sometimes patients perceive pain but they cannot show its signs due to the effects of other drugs. Therefore, the anesthesiologists and clinicians monitor frequently the patients’ vital signs like heart rate, blood pressure, and bispectral index (BIS) and inject the hypnotic drugs such as propofol. Finally, neuromuscular blocking (NMB) drugs NMB drugs facilitate the endotracheal tube insertion.

Besides this category, induction, maintenance and emergency phases are three different steps of intravenous anesthesia. This category is based on the different stages of surgery and defines different tasks for anesthesiologists in each stage. In the induction phase, which is the first phase of anesthesia and happens before surgery starts, the main task of anesthesiologists is decreasing consciousness level of patients to a standard level. This process should be done by an optimal drug infusion and minimum undershoot and overshoot in a standard time frame. As the consciousness level of the patients stabilizes around the desired level, the surgery starts and consequently maintenance phase follows. In this phase, anesthesiologists try to keep the consciousness level constant at the desired level by frequently drug infusions. These frequent drug infusions cover the effect of reducing the impact of previous drugs and eliminate disturbances caused by different steps of surgery. The latest phase is maintenance and happens by stopping drug infusion when the surgery is completely finished.

In the traditional anesthesia, the anesthesiologists defined the amount, rate and time of injection in each phase based on their experience and patients’ vital signs such as heart rate, blood pressure, lacrimation, sweating and papillary dilatation. Although the clinicians’ experience and vital signs contain valuable information about the patients’ state, they do not provide an exact level of the patients’ consciousness. This might increase the probability of incidence of awareness during surgery\textcolor{blue}{\cite{bruhn2006depth}}. Developments in the monitoring of brain activities have given this opportunity to the anesthesiologists to have enough online information about different states of patients during surgery. One of the most common approaches, in order to exact evaluation of anesthesia, is the bispectral (BIS) analysis of the electroencephalogram (EEG) signal. This signal presents electrical activities in the cerebral cortex, which are related to the consciousness level of the patients. This signal is quantified from 0 to 100 and called BIS index and the value 100 denotes full awake or low hypnotic state and as the analgesia or hypnotic drugs inject to the patient, the BIS index decreases. The BIS index 50 defines a desirable consciousness level and a moderate hypnotic state of the patients. This value is acceptable with a tolerance between 60-40\textcolor{blue}{\cite{bailey2006adaptive}}. 

As noted, the developed monitors decrease the probability of clinician’s error and consequently increases patients’ safety in surgery. However, the accuracy of the conventional anesthesia method still strongly depends on the expertise of anesthesiologists to decide about the time and amount of drug infusion and needs to be developed. In the last two decades, closed-loop control systems are presented as a useful solution for designing a sufficient drug delivery system in intravenous anesthesia for example. This approach has some advantage in comparison to the traditional and conventional anesthesia such as preventing too little and much drug infusion, which results in patient awareness and death respectively, anaesthesiologists’ workload reduction, eliminating the possibility of pain-sensing, and decreasing side effects of drugs. In order to design and implement a suitable controller that have all the advantages, control engineers and anesthesiologists are facing several problems. The first problem is inter and intra patient variability that implies physiological parameters of patients are significantly different and change over time and the controller should be always applicable for a variety of patient. Also, the current mathematical models contain nonlinear and time delay terms, which results in some difficulties in designing steps. Another important problem is the presence of measurement devices’ noise and surgical stimulations. These challenges force the controller engineers to design a sufficient controller which is adaptive and robust against the uncertainties. Furthermore, the designed controller should perform two important tasks. The first one is the determination of the depth of anesthesia. This happens in the induction phase and defines that the controller should take best control action (drug infusion rate) to reduce the consciousness level and the pain-sensing state of the patients in the desired time frame with minimum undershoot and overshoot. The second task is control of anesthetic depth, which points to keep the depth of analgesia (DOA) or depth of hypnosis (DOH) constant using BIS output (feedback) and continuous injections. 

Several control strategies have been proposed to develop closed-loop control of anesthesia in recent years. Some studied used classical proportional–integral–derivative (PID) controller and have improved its features to be useful in closed-loop control of anesthesia by applying event base input option\textcolor{blue}{\cite{padula2016inversion, merigo2017event, pawlowski2017event}}, which rejects the noise, and optimize gain tuning and scheduling for each phase of anesthesia\textcolor{blue}{\cite{padula2017optimized}}, which yield less overshoot value and better performance in the induction and maintenance phases. Next developing control structure is the model predictive controller (MPC). Same as PID controller, event base input option improves MPC performance in noise cancelation in anesthesia by reducing drug infusion rate changes and it also makes the process more visible for anesthesiologists\textcolor{blue}{\cite{padula2016inversion, merigo2017event, pawlowski2017event}}. Also, different advanced estimation methods, offset-free and state output correction strategies help the MPC to deal with the inter and intra patient variability challenge\textcolor{blue}{\cite{krieger2014model}}. Both type-1 and type-2 fuzzy logic controllers are also reported in the literature in closed-loop control of anesthesia. These controllers are improved in combination with neural network systems and the genetic algorithm methods and used in the self-organizing fuzzy logic controllers to handle the uncertainties and cancel the noises in anesthesia\textcolor{blue}{\cite{doctor2016type, marrero2017adaptive, khodaei2019adaptive, wang2017improved, tosun2010anesthetic,el2014interval, liu2015genetic}}. Further, other controller strategies such as classical adaptive controllers\textcolor{blue}{\cite{navarro2017fractional, padmanabhan2019optimal}}, nonlinear H-infinity controller\textcolor{blue}{\cite{rigatos2016nonlinear}} and positive state observer controller\textcolor{blue}{\cite{nogueira2016positive}} are employed recently to develop closed-loop control of anesthesia. A comprehensive reviews can be found in\textcolor{blue}{\cite{khodaei2019physiological}}. 

Although the studies have had some significant improvements in closed-loop control of anesthesia, the noted challenges and missions are not still completely addressed. In the current paper, a controller framework is proposed to deal with the current challenges with a sub-optimal control input. “Adaptive multi-critic based neuro-fuzzy controller” (AMCNFC) is developed in\textcolor{blue}{\cite{vatankhah2012adaptive}} and previously tested successfully in the biomedical application by\textcolor{blue}{\cite{vatankhah2016adaptive}} to control endpoint movements of human arms. This control structure is independent of the model parameters and provides the sub-optimal control action in systems with set point or path tracking target. Here, the infusion rate of propofol is determined by this controller to reach the desired depth of hypnosis and stabilize it. It should be noted that the propofol is used as the hypnotic drug since it has a lower negative consequence to other hypnotic drugs\textcolor{blue}{\cite{borgeat1992does}}. In this paper, it is shown that the proposed controller has a significant improvement in closed-loop control of anesthesia, which is the main contribution of this paper, due to the decreasing quantity of the administered drug, improving other performance indexes, and preventing overshoot in BIS response.  

The current paper is formed as follows. The conventional mathematical pharmacokinetic-pharmacodynamics PK/PD model of anesthesia is presented in \textcolor{blue}{Section \ref{Mathematical}} and the controller is explained in \textcolor{blue}{Section \ref{Controller}}. The simulation results are discussed in \textcolor{blue}{Section \ref{simulation}} while a conclusion is posted in \textcolor{blue}{Section \ref{conclusion}}.    

\section{Mathematical modeling} \label{Mathematical}

\begin{figure}
\centering
\includegraphics[width=80mm]{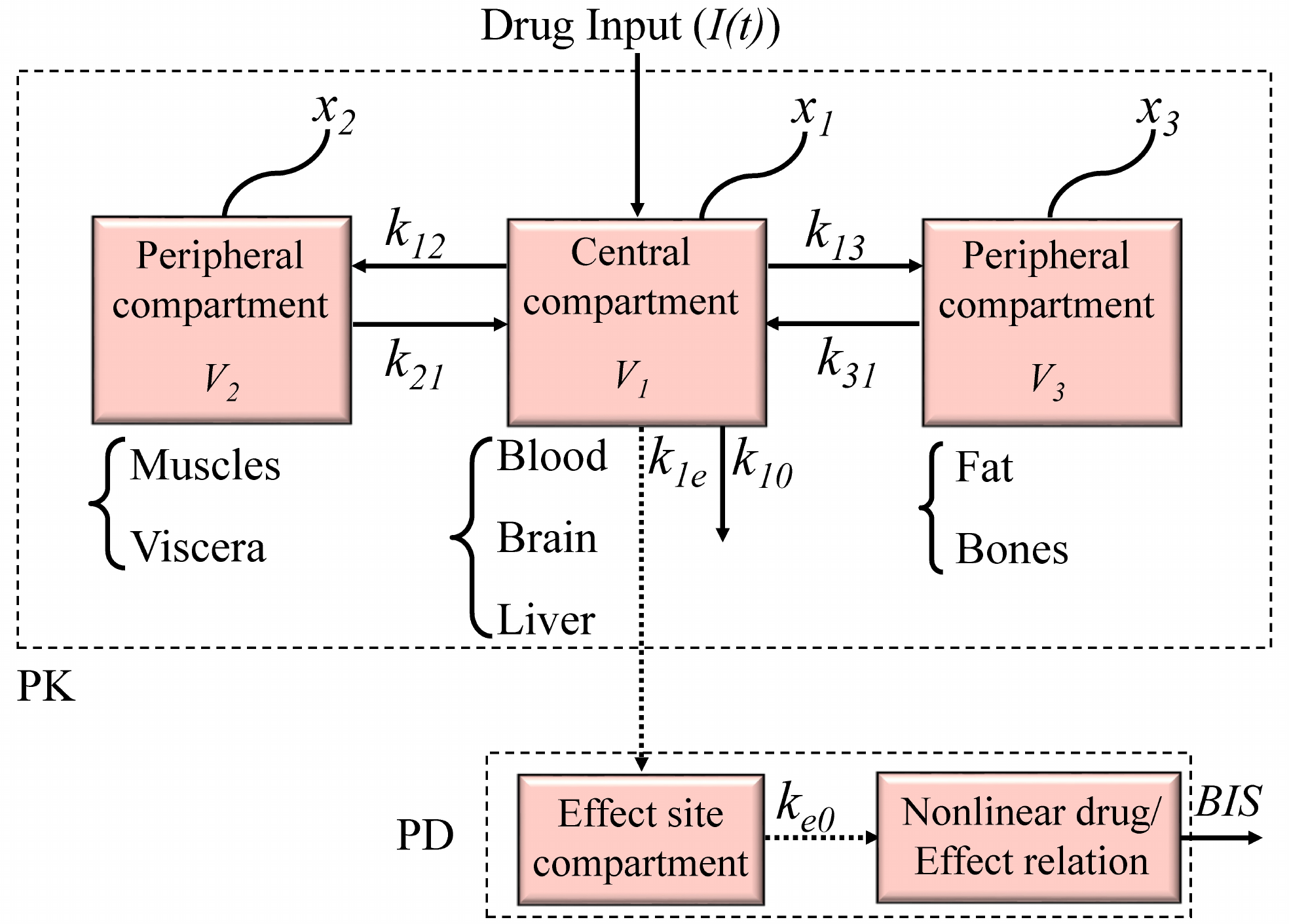}
\caption{The PK/PD compartmental model of anesthesia (Adapted from \textcolor{blue}{\cite{bamdadian2008controlling}}).}
\label{fig2}
\end{figure}

Mathematical modeling of dynamic systems is the first and most significant part in controller design. Complete information and comprehension about the system behavior make the designed controller more efficient. This becomes more complicated for physiological systems due to their complex, nonlinear and variant dynamics. In this section, as can be seen in \textcolor{blue}{Fig. \ref{fig2}}, the conventional PK/PD compartmental model of anesthesia is presented. This is a single input single output model and has three main compartments in the PK part and one effect site compartment in the PD part.  

\subsection{Pharmacokinetic Model} \label{Pharmacokinetic}

When the drugs injected to patients they reach to different parts of the human body. So it is needed to know in what proportion of the injected drugs reach the target. The PK model gives the mathematical model of drug distribution in the human body. As can be seen in \textcolor{blue}{Fig. \ref{fig1}}, this model has 3 important compartments. The compartment $V_1$ is defined as the central compartment and contains important parts such as blood, brain, and liver. The peripheral compartments $V_2$ and $V_3$ represent other human parts, which receive the infused drugs. The compartment $V_2$ includes muscles and viscera, which are well-perfused body tissues and have fast dynamics. On the contrary, the compartment $V_3$ has a slow dynamic since it is consist of fat and bones parts, which are poorly perfused body tissues. This model can be derived by establishing balance equation for drug concentration $x_i (mg)$ in the $i$th compartment as:

\begin{equation}
    \label{eq1}
    {{\dot{x}}_{1}}=\frac{I(t)}{{{V}_{1}}}-{{k}_{12}}{{x}_{1}}-{{k}_{13}}{{x}_{1}}-{{k}_{10}}{{x}_{1}}+{{k}_{21}}{{x}_{2}}+{{k}_{31}}{{x}_{3}}
\end{equation}

\begin{equation}
    \label{eq2}
    {{\dot{x}}_{2}}={{k}_{12}}{{x}_{1}}-{{k}_{21}}{{x}_{2}}
\end{equation}

\begin{equation}
    \label{eq3}
    {{\dot{x}}_{3}}={{k}_{13}}{{x}_{1}}-{{k}_{31}}{{x}_{3}}
\end{equation}
where constants $k_{ij}(min^{-1})$ denote the transfer rate of the drug from compartment $i$ to compartment $j$ and are different for each patient, the constant $k_{10}$ shows the rate of the drug metabolism. In Eq. \eqref{eq1}, $I(t)(mg/min)$ is the drug infusion rate in the central compartment. These equations can be rewritten in state space format as below\textcolor{blue}{\cite{bailey2005drug}}:

\begin{align}
\begin{split}
    \label{eq4}
\left[ \begin{matrix}
   {{{\dot{x}}}_{1}}  \\
   {{{\dot{x}}}_{2}}  \\
   {{{\dot{x}}}_{3}}  \\
\end{matrix} \right]=&\left( \begin{matrix}
   -\left( {{k}_{12}}+{{k}_{13}}+{{k}_{10}} \right) & {{k}_{21}} & {{k}_{31}}  \\
   {{k}_{12}} & -{{k}_{21}} & 0  \\
   {{k}_{13}} & 0 & -{{k}_{31}}  \\
\end{matrix} \right)\left[ \begin{matrix}
   {{x}_{1}}  \\
   {{x}_{2}}  \\
   {{x}_{3}}  \\
\end{matrix} \right] \\
&+\left[ \begin{matrix}
   \frac{1}{{{V}_{1}}}  \\
   0  \\
   0  \\
\end{matrix} \right]I(t)
\end{split}
\end{align}

\begin{equation}
    \label{eq5}
    {{C}_{p}}=\left[ \begin{matrix}
   1 & 0 & 0  \\
\end{matrix} \right]\left[ \begin{matrix}
   {{x}_{1}}  \\
   {{x}_{2}}  \\
   {{x}_{3}}  \\
\end{matrix} \right]
\end{equation}
where $C_P$ is the amount of the drug in blood plasma and $k_{ij}$ can be obtained as below for the propofol\textcolor{blue}{\cite{schnider1998influence}}:

\begin{equation}
    \label{eq6}
    \left[ \begin{matrix}
   {{k}_{10}}  \\
   {{k}_{12}}  \\
   {{k}_{21}}  \\
   \begin{matrix}
   {{k}_{31}}  \\
   {{k}_{13}}  \\
\end{matrix}  \\
\end{matrix} \right]=\left( \begin{matrix}
   \begin{matrix}
   \frac{1}{{{V}_{1}}} & 0 & 0  \\
\end{matrix}  \\
   \begin{matrix}
   0 & \frac{1}{{{V}_{1}}} & 0  \\
\end{matrix}  \\
   \begin{matrix}
   0 & 0 & \frac{1}{{{V}_{1}}}  \\
\end{matrix}  \\
   \begin{matrix}
   \begin{matrix}
   0 & \frac{1}{{{V}_{1}}} & 0  \\
\end{matrix}  \\
   \begin{matrix}
   0 & 0 & \frac{1}{{{V}_{1}}}  \\
\end{matrix}  \\
\end{matrix}  \\
\end{matrix} \right)\left[ \begin{matrix}
   C{{l}_{1}}  \\
   C{{l}_{2}}  \\
   C{{l}_{3}}  \\
\end{matrix} \right]
\end{equation}
where $V_i(l)$ is the volume of compartment $i$ and can be calculated as:

\begin{equation}
    \label{eq7}
\left[ \begin{matrix}
   {{V}_{1}}  \\
   {{V}_{2}}  \\
   {{V}_{3}}  \\
\end{matrix} \right]=\left[ \begin{matrix}
   0  \\
   -0.391  \\
   0  \\
\end{matrix} \right]\left[ Age \right]+\left[ \begin{matrix}
   4.27  \\
   39.623  \\
   238  \\
\end{matrix} \right]
\end{equation}
and age is in the year. In Eq. \eqref{eq6}, $Cl_i(m/l)$ can be calculated as:

\begin{align}
\begin{split}
    \label{eq8}
    \left[ \begin{matrix}
   C{{l}_{1}}  \\
   C{{l}_{2}}  \\
   C{{l}_{3}}  \\
\end{matrix} \right]=&\left( \begin{matrix}
   \begin{matrix}
   0.0456  \\
   0  \\
   0  \\
\end{matrix} & \begin{matrix}
   0.0264  \\
   0  \\
   0  \\
\end{matrix} & \begin{matrix}
   -0.0681  \\
   0  \\
   0  \\
\end{matrix} & \begin{matrix}
   0  \\
   -0.024  \\
   0  \\
\end{matrix}  \\
\end{matrix} \right) \times \\
&\left[ \begin{matrix}
   Weight  \\
   Height  \\
   lbm  \\
   Age  \\
\end{matrix} \right]\,+\left[ \begin{matrix}
   -2.271  \\
   0.018  \\
   0.836  \\
\end{matrix} \right]
\end{split}
\end{align}
where weight and height are in $kg$ and $cm$, respectively. In Eq. \eqref{eq8}, lbm is the lean body mass and its formula is different for males and females. It can be calculated as\textcolor{blue}{\cite{hallynck1981should}}:

\begin{equation}
    \label{eq9}
\left[ \begin{matrix}
   lb{{m}_{m}}  \\
   lb{{m}_{f}}  \\
\end{matrix} \right]=\left( \begin{matrix}
   m & 0  \\
   0 & f  \\
\end{matrix} \right)\left( \begin{matrix}
   1.1 & -128  \\
   1.07 & -148  \\
\end{matrix} \right)\left[ \begin{matrix}
   Weight  \\
   {{\left( \frac{Weight}{Height} \right)}^{2}}  \\
\end{matrix} \right]
\end{equation}

In this equation when the patients are male the $m$ and $f$ parameters are equal to one and zero and when they are female these values are reversed.  

\subsection{Pharmacodynamics Model}\label{Pharmacodynamics}

The pharmacodynamics model points to express what happens when the infused drug is in the target compartment. This model quantifies the effect of the drug concentration on the target compartment. Effect site compartment and Hill equation are two parts of this model. The effect site compartment is first introduced by\textcolor{blue}{\cite{sheiner1979simultaneous}} and its volume size is negligible in comparison with other compartments\textcolor{blue}{\cite{shafer1998principles}}. Same as the PK part, the drug concentration equation of this compartment can be derived as:

\begin{equation}
    \label{eq10}
    {{\dot{C}}_{e}}={{\dot{x}}_{e}}={{k}_{1e}}{{x}_{1}}-{{k}_{e0}}{{x}_{e}}
\end{equation}
where $x_e$ denotes drug concentration in the effect site compartment and $k_{1e}$ and $k_{e0}$ are transfer rates of the drug from the central to the effect site compartment and vice versa. These parameters are usually assumed to be the same and are equal to 0.456 $min^{-1}$ for propofol\textcolor{blue}{\cite{schnider1998influence}}. So, the Eq.10 can be rewritten as:  

\begin{equation}
    \label{eq11}
    {{\dot{C}}_{e}}={{k}_{e0}}\left( {{C}_{p}}-{{C}_{e}} \right)
\end{equation}

The second and most important part of the PD model is the Hill equation (sigmoid function). This equation defines the relation between drug concentration in the effect site compartment and the BIS index. The nonlinear Hill equation is given as:     

\begin{equation}
    \label{eq12}
    E={{E}_{0}}-{{E}_{\max }}\frac{C_{e}^{\gamma }}{C_{e}^{\gamma }+EC_{50}^{\gamma }}
\end{equation}
where $E_0$ indicates the baseline value (obtained from the awake state without propofol), which is typically set to 100; $E_{max}$ is the maximum effect achievable by the drug infusion; $EC_{50}$ denotes the drug concentration at half maximal effect, which represents the patient’s sensitivity to the drug and should be measured experimentally and $\gamma$ determines the slope of the sigmoid curve (i.e., the receptiveness of the patient to the drug)\textcolor{blue}{\cite{nino2009epsac}}. As noted, $E$ in Eq. 12 gives the BIS index in the range of [0, 100]. The parameters of the Hill equation should be obtained experimentally and are different for each patient. Here, a set of 12 patients’ parameters for propofol infusion are presented in \textcolor{blue}{Table \ref{table1}}. This data base is previously used in literature to simulate the PK/PD model and closed-loop control of anesthesia. In \textcolor{blue}{Table \ref{table1}}, the thirteenth patient named nominal patient and its parameters is the average of the 12 patients. 

\begin{table*}
\centering
\caption{Numerical values of the Hill equation’s parameter\textcolor{blue}{\cite{nino2009epsac}}}
\label{table1}
\begin{tabular}{ccccccccc}
\hline \hline
Patient $\#$ & Age & Length (cm) & Weight (kg) & Gender & $EC_{50}$ & $E_0$ & $E_{max}$ & $\gamma$ \\
\hline \hline
1 & 40 & 163 & 54 & F & 6.33 & 98.8 & 94.1 & 2.24 \\
2 & 36 & 163 & 50 & F & 6.76 & 98.6 & 86 & 4.29 \\
3 & 28 & 164 & 52 & F & 8.44 & 91.2 & 80.7 & 4.1 \\
4 & 50 & 163 & 83 & F & 6.44 & 95.9 & 102 & 2.18 \\
5 & 28 & 164 & 60 & M & 4.93 & 94.7 & 85.3 & 2.46 \\
6 & 43 & 163 & 59 & F & 12.1 & 90.2 & 147 & 2.42 \\
7 & 37 & 187 & 75 & M & 8.02 & 92 & 104 & 2.10 \\
8 & 38 & 174 & 80 & F & 6.56 & 95.5 & 76.4 & 4.12 \\
9 & 41 & 170 & 70 & F & 6.15 & 89.2 & 63.8 & 6.89 \\
10 & 37 & 167 & 58 & F & 13.7 & 83.1 & 151 & 1.65 \\
11 & 42 & 179 & 78 & M & 4.82 & 91.8 & 77.9 & 1.85 \\
12 & 34 & 172 & 58 & F & 4.95 & 96.2 & 90.8 & 1.84 \\
13 & 38 & 169 & 65 & F & 7.42 & 93.1 & 96.6 & 3 \\ \hline \hline
\end{tabular}
\end{table*}

\section{Controller design}\label{Controller}

\begin{figure}
\centering
\includegraphics[width=80mm]{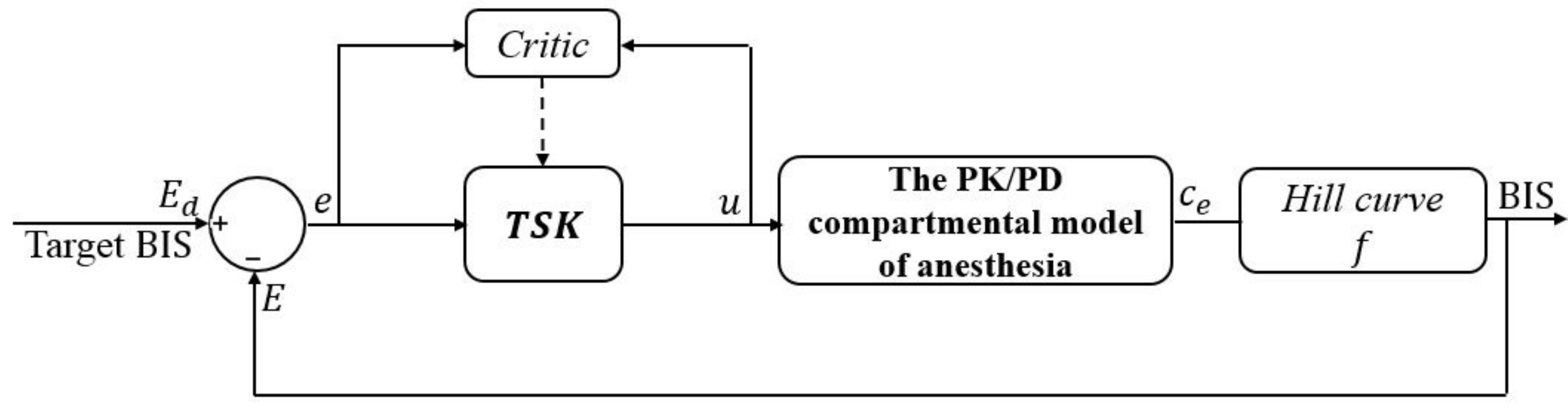}
\caption{Schematic of the controller structure.}
\label{fig3}
\end{figure}

In order to solve the common challenges of the closed-loop anesthesia control such as intra and inter patient variability and the effects of noise, we utilized a kind of adaptive critic-based neuro-fuzzy controller. This controller is model independent and can optimize the amount of drug infusion rate. The main goal of the controller is to minimize the overshoot values during the induction and disturbance rejection phases. In \textcolor{blue}{Fig. \ref{fig3}} a schematic view of this closed-loop control is shown. In the following parts, we explain different parts of the controller.

\subsection{Neuro-fuzzy controller}\label{Neuro-fuzzy}

A fuzzy system is a method of representing the input-output relationship of the systems which is inferred by experimental data. Its structure is based on IF-THEN rules. The inputs and outputs of the IF-THEN rules are represented by using continues membership functions\textcolor{blue}{\cite{soliman2011modern}}. Fuzzy systems can be used as the controller due to its ability to provide any desired non-linear mapping between inputs and outputs. For our specific application, we use a single-input-single-output TSK-type fuzzy system which is the combination of IF-THEN rules as follows:

\begin{equation}
    \label{eq13}
    \begin{matrix}
   IF & \begin{matrix}
   (e & is & {{F}_{i}})  \\
\end{matrix} & THEN & {{C}_{i}}(e)={{a}_{i}}e+{{b}_{i}}  \\
\end{matrix}
\end{equation}
where $e=(E_{d}-E)/100$ is the error between the measured and the desired BIS value, $i\in\{1,N\}$ represents the rule number, $F_i$ is related to its membership function, $C_i$ denotes the output of rule $i$ in which $a_i$ and $b_i$ express the constant coefficients. If the triggering value of each rule is $\mu_i (e)$, then the net output of the TSK system is

\begin{equation}
    \label{eq14}
    u=\sum\limits_{i=1}^{N}{{{\mu }_{i}}(e){{C}_{i}}(e)}.
\end{equation}

\begin{figure}
\centering
\includegraphics[width=80mm]{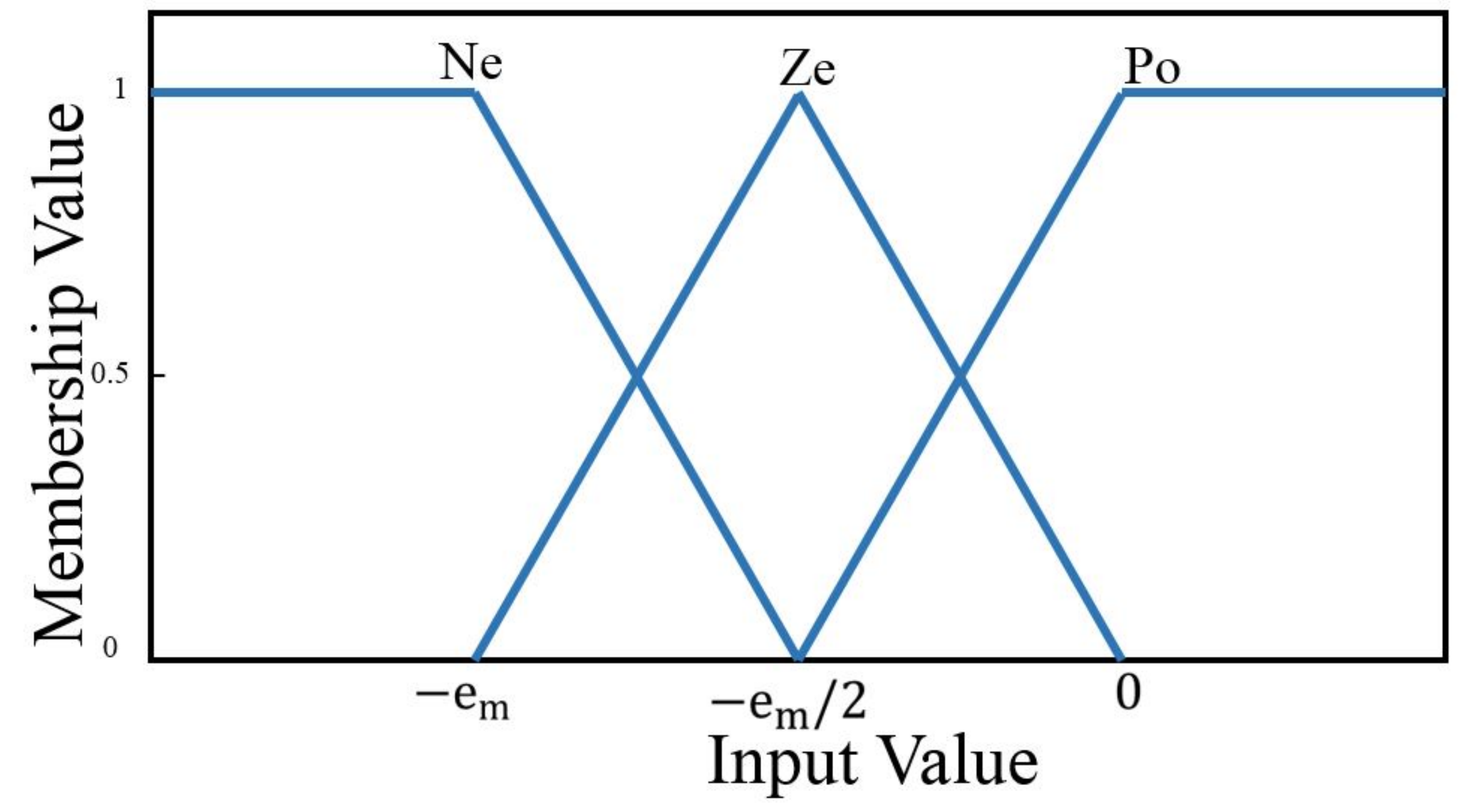}
\caption{Membership functions for TSK controller.}
\label{fig4}
\end{figure}

\begin{figure}
\centering
\includegraphics[width=80mm]{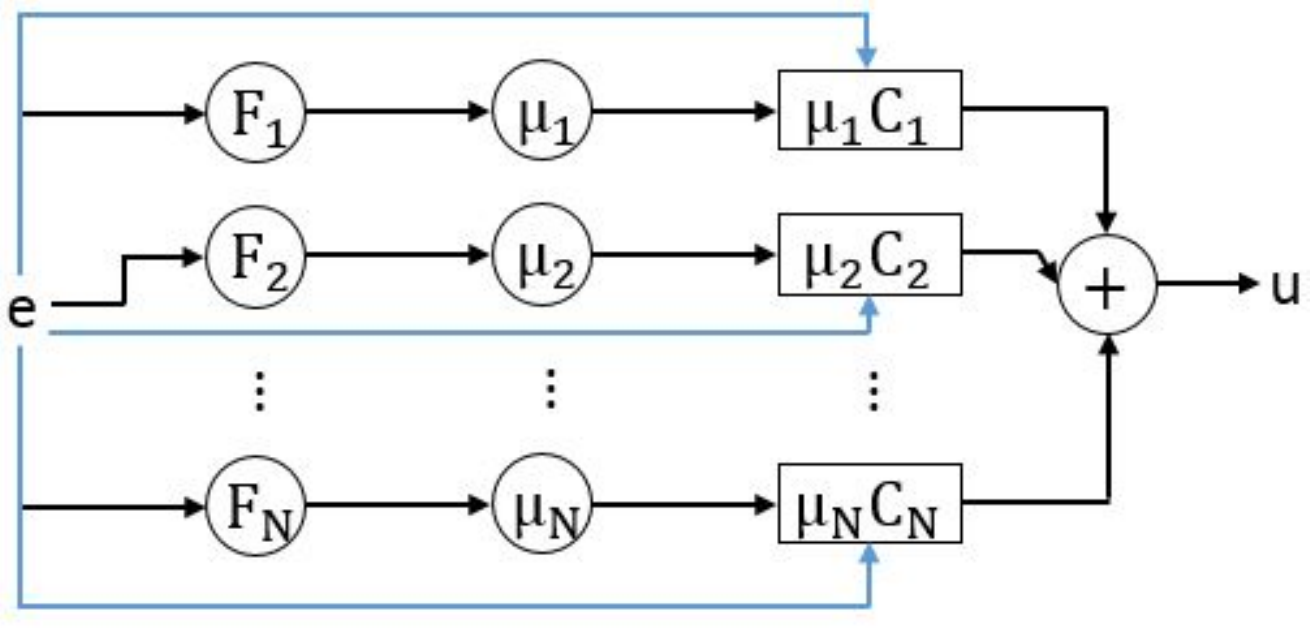}
\caption{The TSK structure.}
\label{fig5}
\end{figure}

For the proposed controller, as shown in \textcolor{blue}{Fig. \ref{fig4}}, we consider three triangular membership functions ($N=3$) named as Negative ($Ne$), Positive ($Po$) and Zero ($Ze$) for the input of the TSK fuzzy controller ($e$). This results in three rules for the TSK controller of anesthesia, which is shown in \textcolor{blue}{Fig. \ref{fig5}}. By using \eqref{eq14}, the proposed TSK fuzzy controller can be represented in the vector form as follows 

\begin{equation}
    \label{eq15}
    u=\bar{\mu }\alpha X
\end{equation}

\begin{equation}
    \label{eq16}
    \bar{\mu }=\left[ \begin{matrix}
   {{\mu }_{1}}(e) & {{\mu }_{2}}(e) & {{\mu }_{3}}(e)  \\
\end{matrix} \right]
\end{equation}

\begin{equation}
    \label{eq17}
    \alpha =\left( \begin{matrix}
   \begin{matrix}
   {{a}_{1}} & {{b}_{1}}  \\
\end{matrix}  \\
   \begin{matrix}
   {{a}_{2}} & {{b}_{2}}  \\
\end{matrix}  \\
   \begin{matrix}
   {{a}_{3}} & {{b}_{3}}  \\
\end{matrix}  \\
\end{matrix} \right)
\end{equation}

\begin{equation}
    \label{eq18}
    X=\left[ \begin{matrix}
   e  \\
   1  \\
\end{matrix} \right]
\end{equation}

\subsection{The Critic}\label{Critic}

As shown in \textcolor{blue}{Fig. \ref{fig3}}, the parameters of the neuro-fuzzy controller ($a_i$ and $b_i$) should be updated by using a reinforcement signal, $r$. This scaler signal is produced by a critic agent and is determined by evaluating the output of the system. The idea is to adapt the tunable parameters of the TSK fuzzy controller by using this performance criteria. The goal is to force the controller to minimize the critic signal ($r$) so that it reaches zero value\textcolor{blue}{\cite{hatami2014design}} and at the same time minimize the control input ($u$). Therefore, the cost function is defined as:

\begin{equation}
    \label{eq19}
    J=\frac{1}{2}k{{r}^{2}}+\frac{1}{2}K{{u}^{2}}
\end{equation}
in which $k$ and $K$ are constant positive values and

\begin{equation}
    \label{eq20}
    r=e
\end{equation}

As seen, the cost function is composed of two terms. The first terms are for minimizing the reinforcement signal which is equal to the BIS error value in this application,  while the second term is used for minimizing the control input, which is the drug infusion rate. 

\subsection{Emotional Learning}\label{Emotional}

Emotional learning is referred to the method by which the controller's parameters are updated by using the reinforcement signal and the control input signal. We use the steepest descent algorithm to achieve this goal. This procedure is as follows:

\begin{equation}
    \label{eq21}
    \dot{\alpha }=-\eta \frac{\partial J}{\partial \alpha }
\end{equation}
where $\eta$ is the learning rate of the controller. Using partial differentiation, it will be simplified as follows

\begin{equation}
    \label{eq22}
    \dot{\alpha }=-\eta \left( \frac{\partial J}{\partial r}\frac{\partial r}{\partial E}\frac{\partial E}{\partial u}+\frac{\partial J}{\partial u} \right)\frac{\partial u}{\partial \alpha }
\end{equation}

In this application, the system is the minimum phase and therefore, $\frac{\partial E}{\partial u}\approx 1$, then

\begin{equation}
    \label{eq23}
    \dot{\alpha}=-\eta \left(-kr+Ku\right) \bar{\mu}^{T}X^{T}
\end{equation}
which is the dynamic of the TSK fuzzy controller’s parameters. The initial value   can be selected randomly.

\section{Simulation}\label{simulation}

The proposed controller is designed based on the nominal patient information and in this section its ability in closed-loop control of the PK/PD compartment model is assessed by simulation results. These simulation results are presented in both induction and maintenance phases for the $12$ patients. In the designed controller, the TSK parameters are needed to be selected for the first treatment. These parameters are selected randomly between $-2$ and $2$. The learning rate is also selected $2$ and the $e_m$ in \textcolor{blue}{Fig. \ref{fig4}} is chosen $0.5$. Since the current infusion pumps have some limitations range and the infusion rate cannot be a negative value, the infusion rate is limited in the range of $[0, 50]$ \textcolor{blue}{\cite{nascu2017modeling}}. Also, the sampling rate is selected $1$ second.  

\subsection{Induction phase}\label{Induction}

\begin{figure}
\centering
\includegraphics[width=80mm]{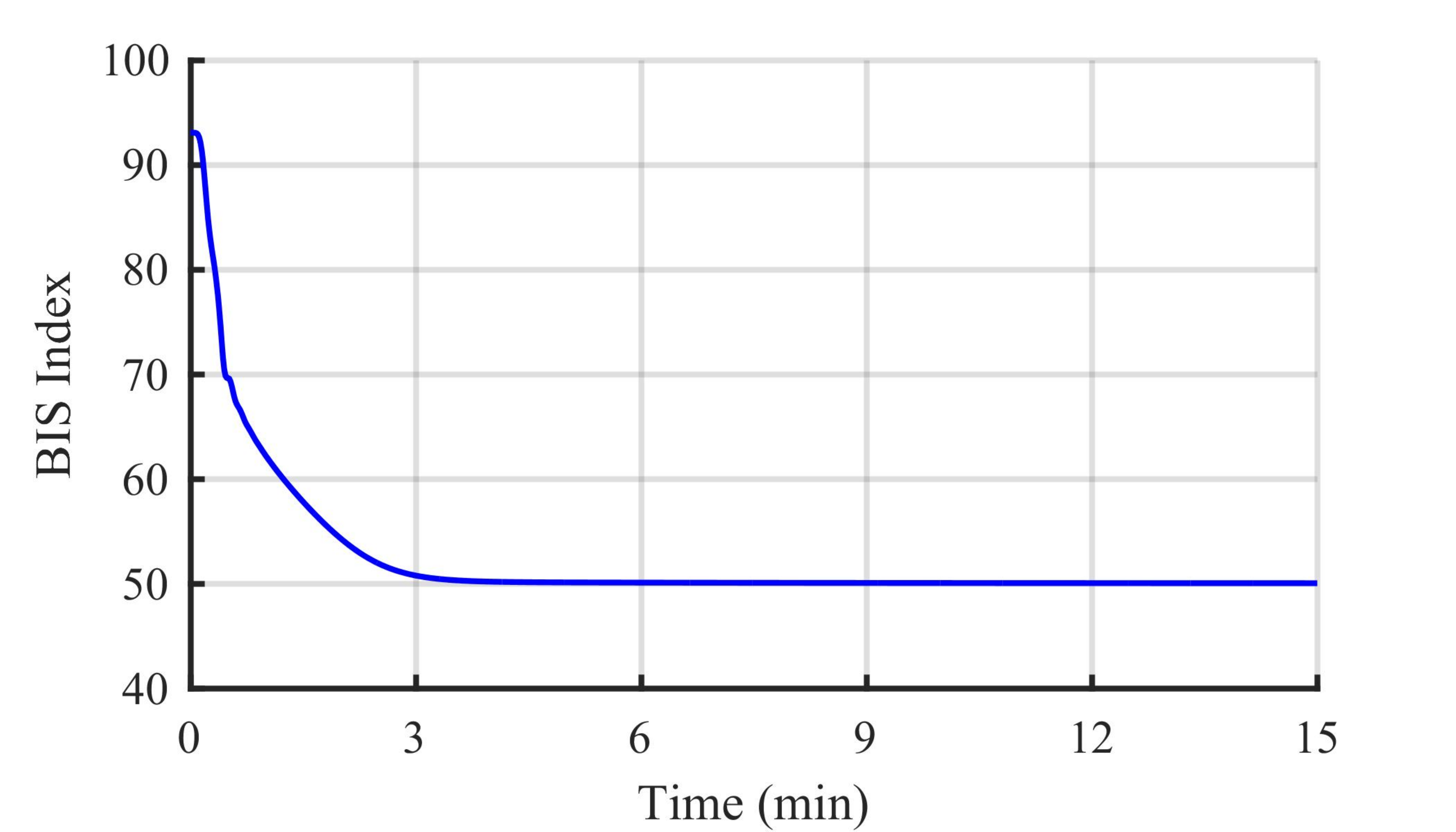}
\caption{BIS output for the nominal patient in the induction phase.}
\label{fig6}
\end{figure}

\begin{figure}
\centering
\includegraphics[width=80mm]{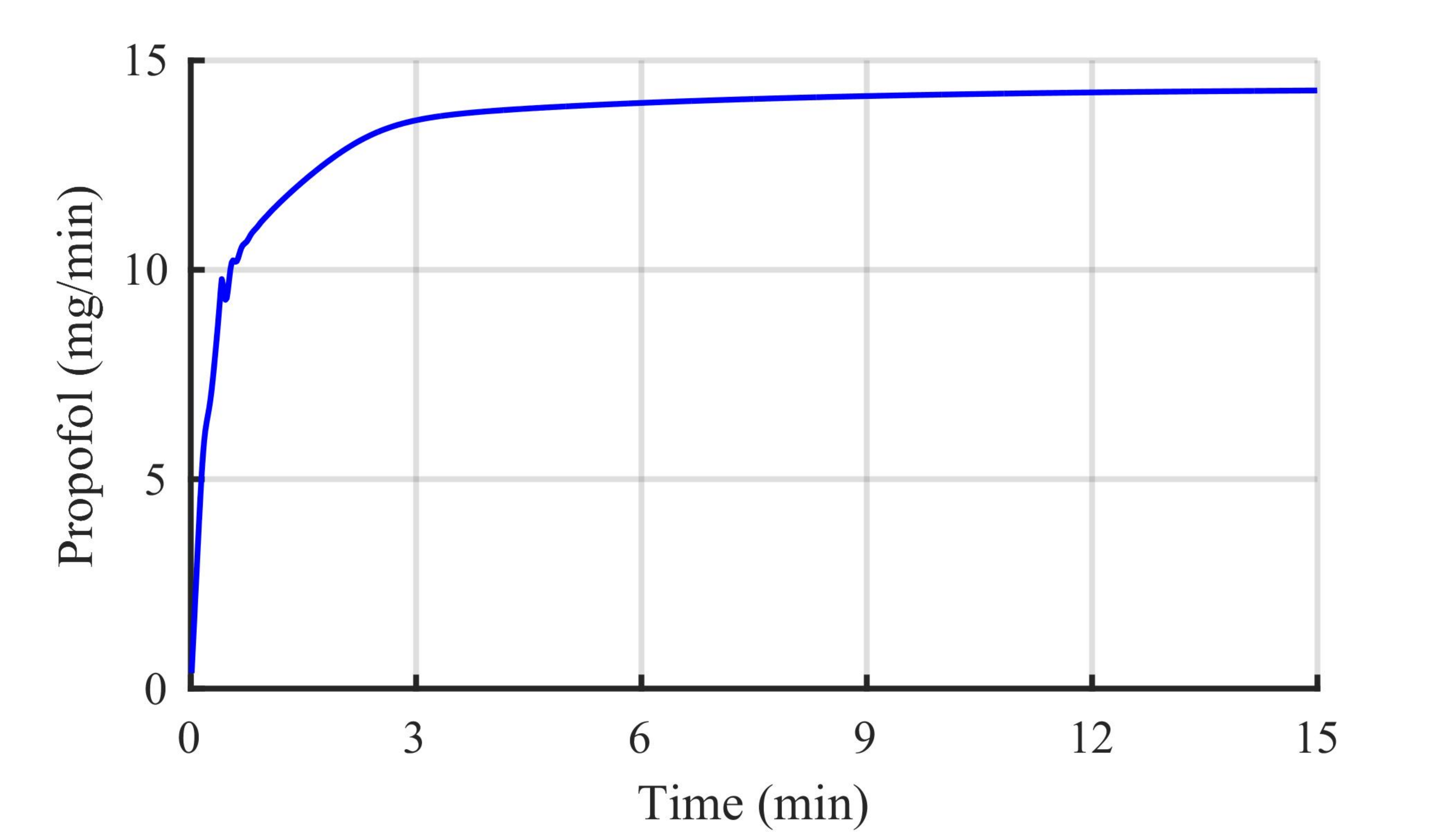}
\caption{Drug Infusion for the nominal patient in the induction phase.}
\label{fig7}
\end{figure}

In this section, the controller is evaluated in the induction phase. In this phase, the controller should provide an optimal drug infusion rate to reduce the BIS value to $50$, which is the desired DOH, in an appropriate time range with no overdose. The settling time is a significant factor in this process and some parameters are considerable on the settling time definition. As the learning rate increases the settling time decreases and this makes more drug infusion. However, too little settling time is not acceptable by anesthesiologists due to the increase in drug consumption and their other side effects. Actually, infusion of a huge drug amount in a short settling time shocks the human body and might harm other physiological organs of the patients\textcolor{blue}{\cite{das2014control}}. However, a long settling time can be obtained by a lower infusion rate. The definition of this time needs a trade-off between the side effects and the drug infusion rate. In the current surgical operations, this time is reported 15 minutes\textcolor{blue}{\cite{nascu2017modeling}} while shorter time is defined in the literature for simulation studies. In this paper this time is selected to be in the range of $2-4$ minutes. The controller is applied to the PK/PD model of the nominal patient and its potency in control of the BIS value is shown in \textcolor{blue}{Fig. \ref{fig6}} while \textcolor{blue}{Fig. \ref{fig7}} represents the control action. \textcolor{blue}{Fig. \ref{fig6}} shows the desired DOH is obtained in the defined settling time. It also indicates that the provided sub-optimal control action prevents any overshoot or undershoot close to the desired BIS value.   

\begin{figure}
\centering
\includegraphics[width=80mm]{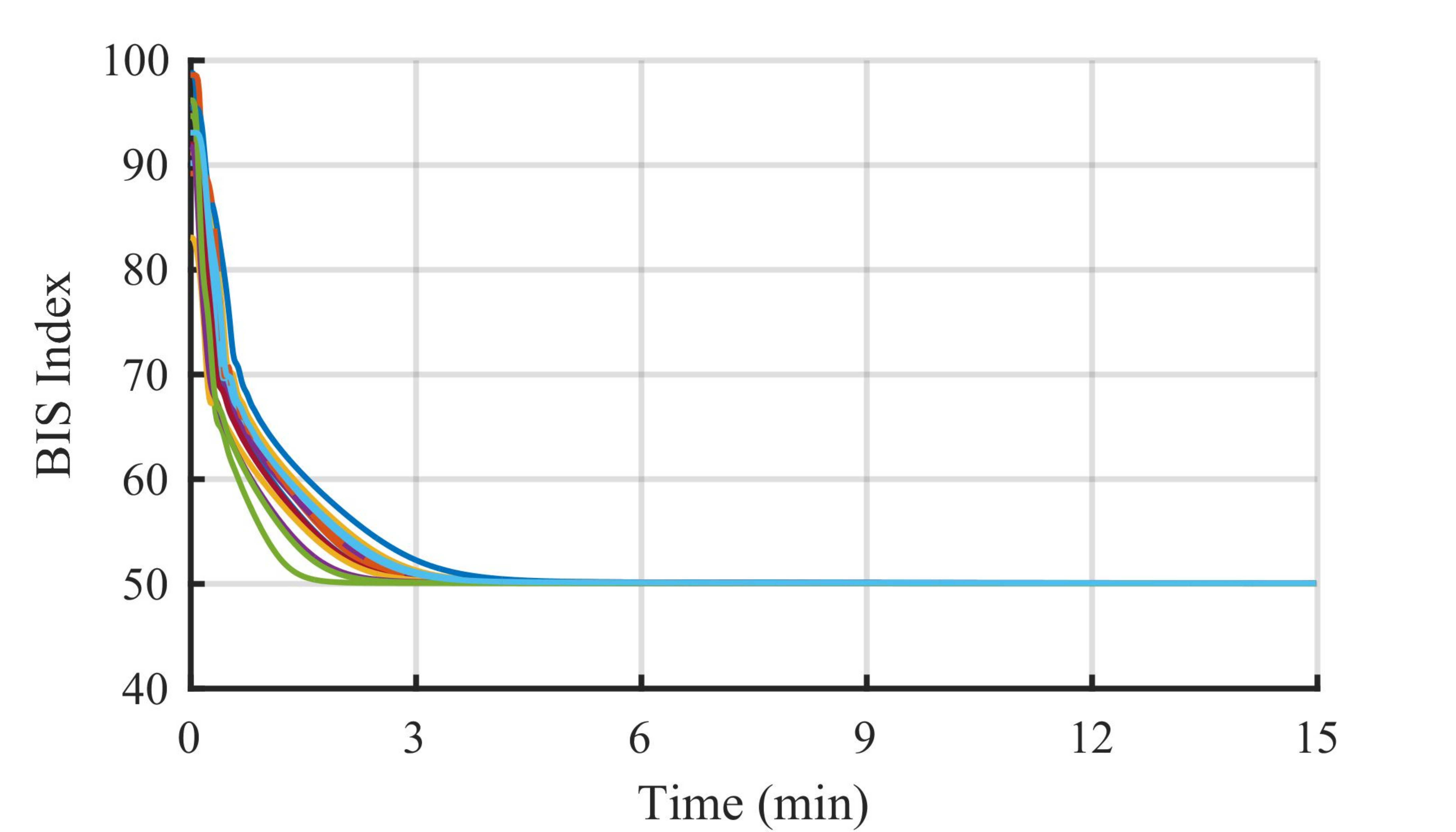}
\caption{BIS output for all patients in the induction phase (patient 9: dash line).}
\label{fig8}
\end{figure}

\begin{figure}
\centering
\includegraphics[width=80mm]{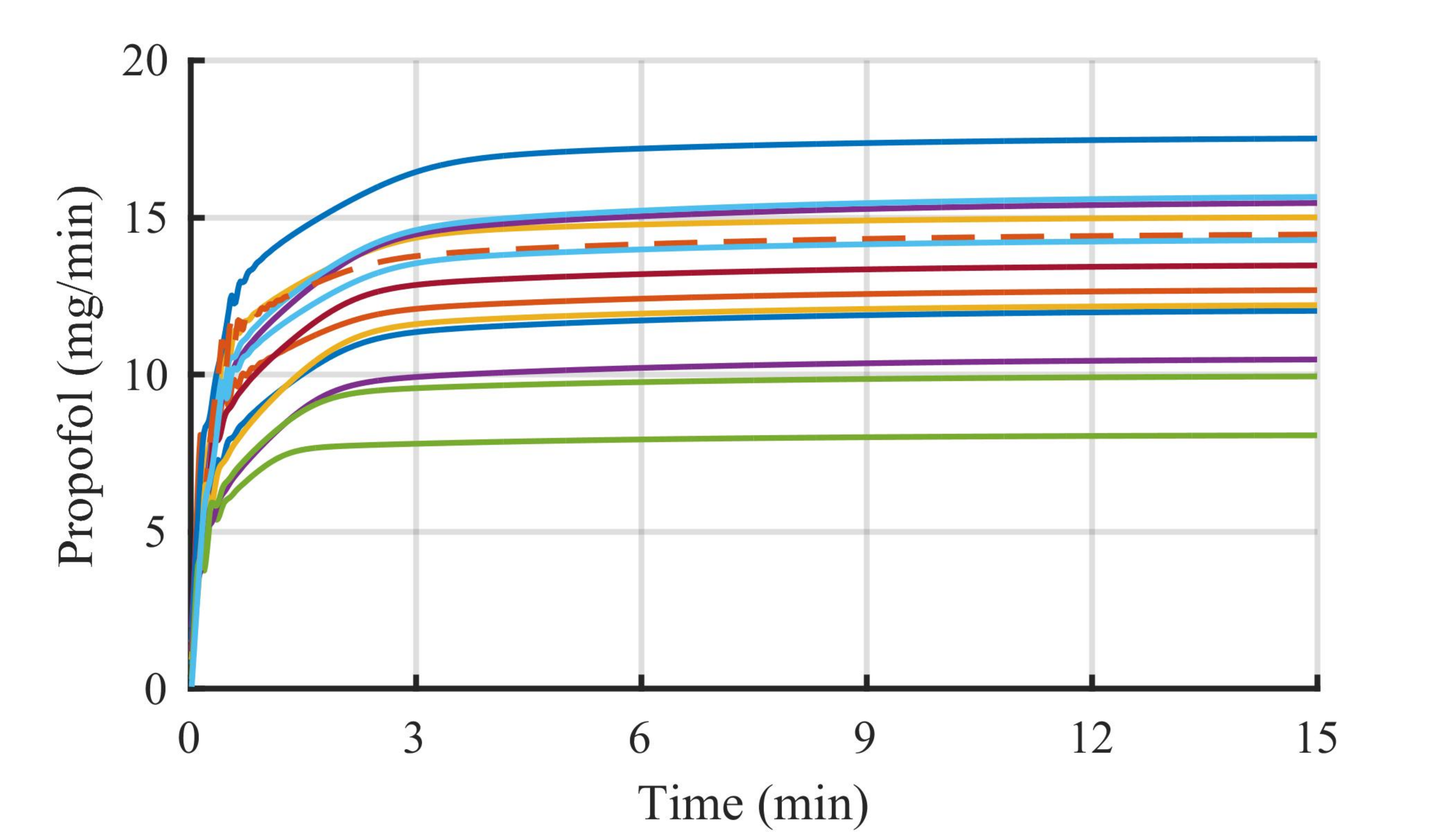}
\caption{Drug Infusion for all patients in the induction phase (patient 9: dash line).}
\label{fig9}
\end{figure}

\begin{figure}
\centering
\includegraphics[width=80mm]{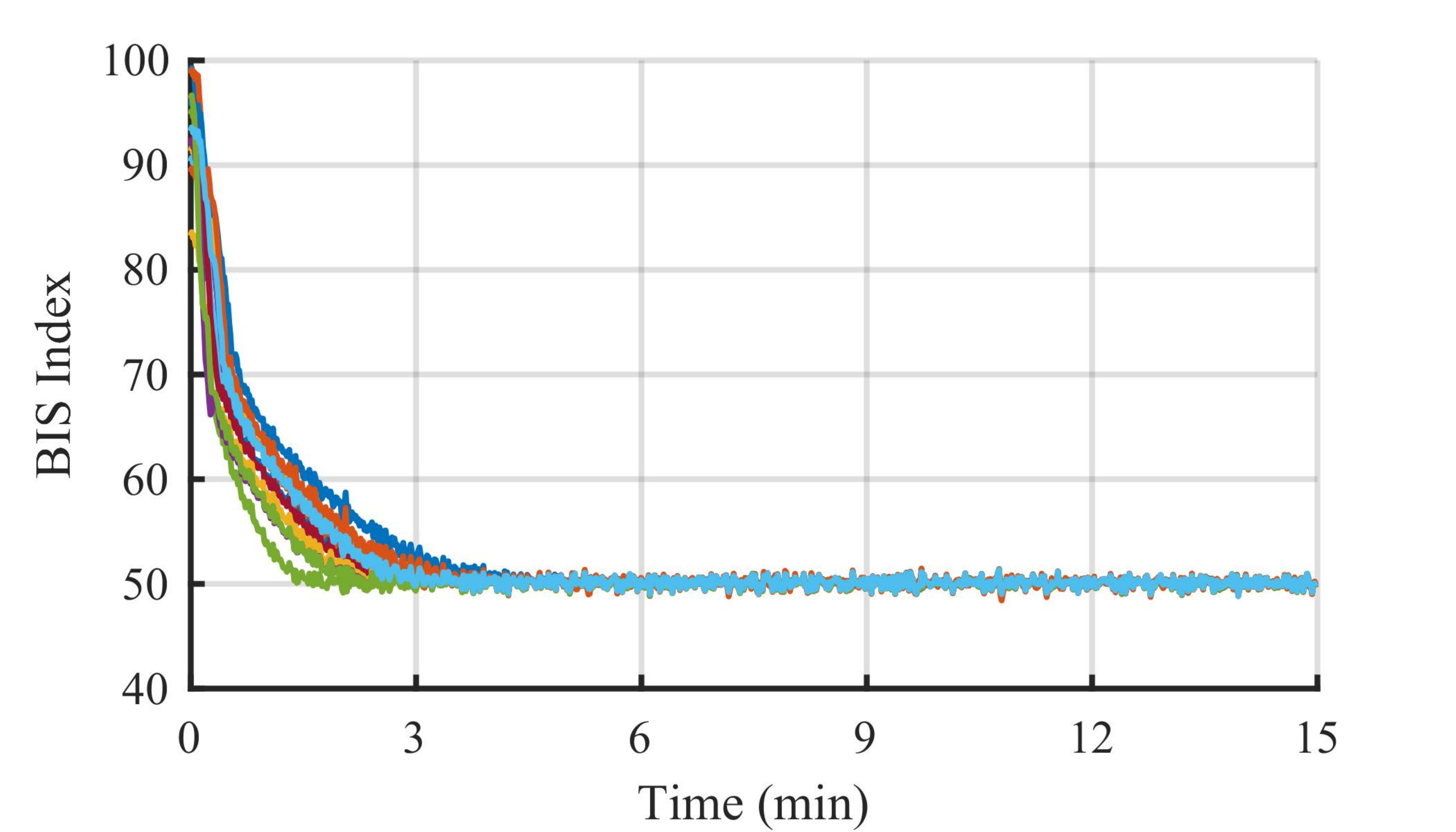}
\caption{Noise effect on BIS output for all patients in the induction phase.}
\label{fig10}
\end{figure}

\begin{figure}
\centering
\includegraphics[width=80mm]{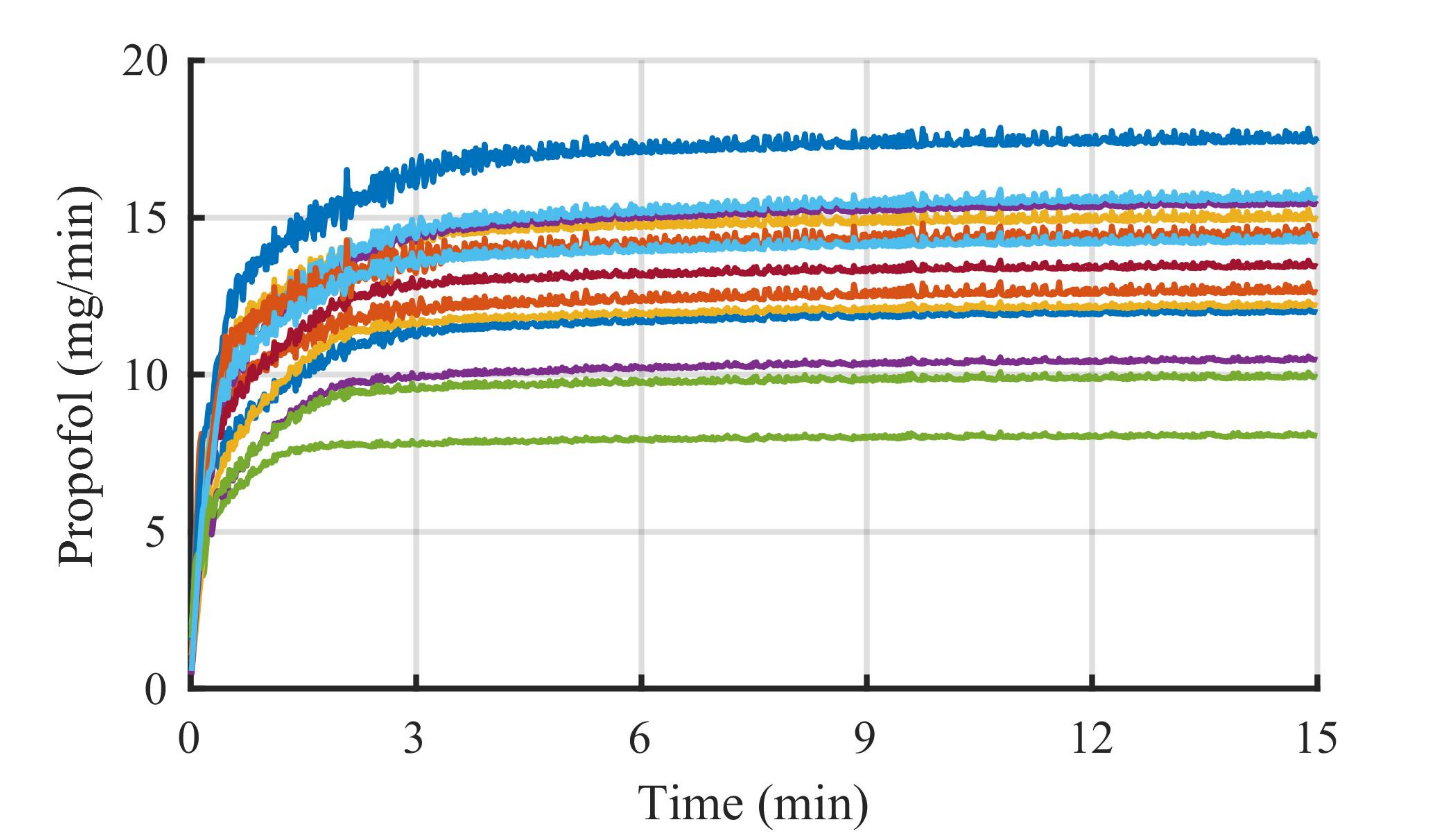}
\caption{Drug Infusion for all patients in the induction phase in the presence of noise.}
\label{fig11}
\end{figure}

The next test is evaluating the proposed controller in the presence of inter and intra patient variability challenge. This implies that the controller should be applied to the PK/PD model of other patients in \textcolor{blue}{Table \ref{table1}}. It should be noted that the patient 9 is defined as the most sensitive and challengeable patient in\textcolor{blue}{\cite{nascu2017modeling}}.  \textcolor{blue}{Fig. \ref{fig8}} and \textcolor{blue}{Fig. \ref{fig9}} indicate the controller can deal with the challenge successfully. As shown, for different patients the designed controller provides different control action while its performance and the system output does not change significantly due to lack of overshoot and long settling time in the responses. To evaluate the controller performance against the measurement devices’ noise, a band-limited white noise is applied to the system output (BIS values). Since the controller is independence from the model and optimizes the control action online, as can be seen in \textcolor{blue}{Fig. \ref{fig10}} and \textcolor{blue}{Fig. \ref{fig11}}, the controller is robust against the noises.  

\begin{table*}
\centering
\caption{Performance indexes values of the controllers}
\label{table2}
\begin{tabular}{cccc}
\hline \hline
Index & PID controller\textcolor{blue}{\cite{padula2017optimized}} & Event-based PID controller\textcolor{blue}{\cite{merigo2017event}} & AMCNFC \\ \hline \hline
IAE & 3499 (for Worst-case IAE) & 3932 (for Nominal patient) & \begin{tabular}[c]{@{}c@{}}2918.9 (for Worst-case)\\ 2215.9 (for Nominal patient)\end{tabular} \\
MDPE {[}\%{]} & 0.46± 1.16 & 0.61± 0.51 & 0.3± 0.09 \\
MDAPE {[}\%{]} & 1.21± 1.33 & 1.44± 0.71 & 0.3± 0.09 \\
WOBBLE {[}\%{]} & 1.21± 0.56 & 2.34± 0.95 & 0.1± 0.05 \\
TV & 60.97± 2.53 & 4.3± 0.74 & 13.8± 3.88 \\
Q {[}mg{]} & 317.71± 51.84 & 308.12± 50.25 & 120.67± 24.38 \\
\hline \hline
\end{tabular}
\end{table*}

\subsection{Maintenance phase}\label{Maintenance}

\begin{figure}
\centering
\includegraphics[width=80mm]{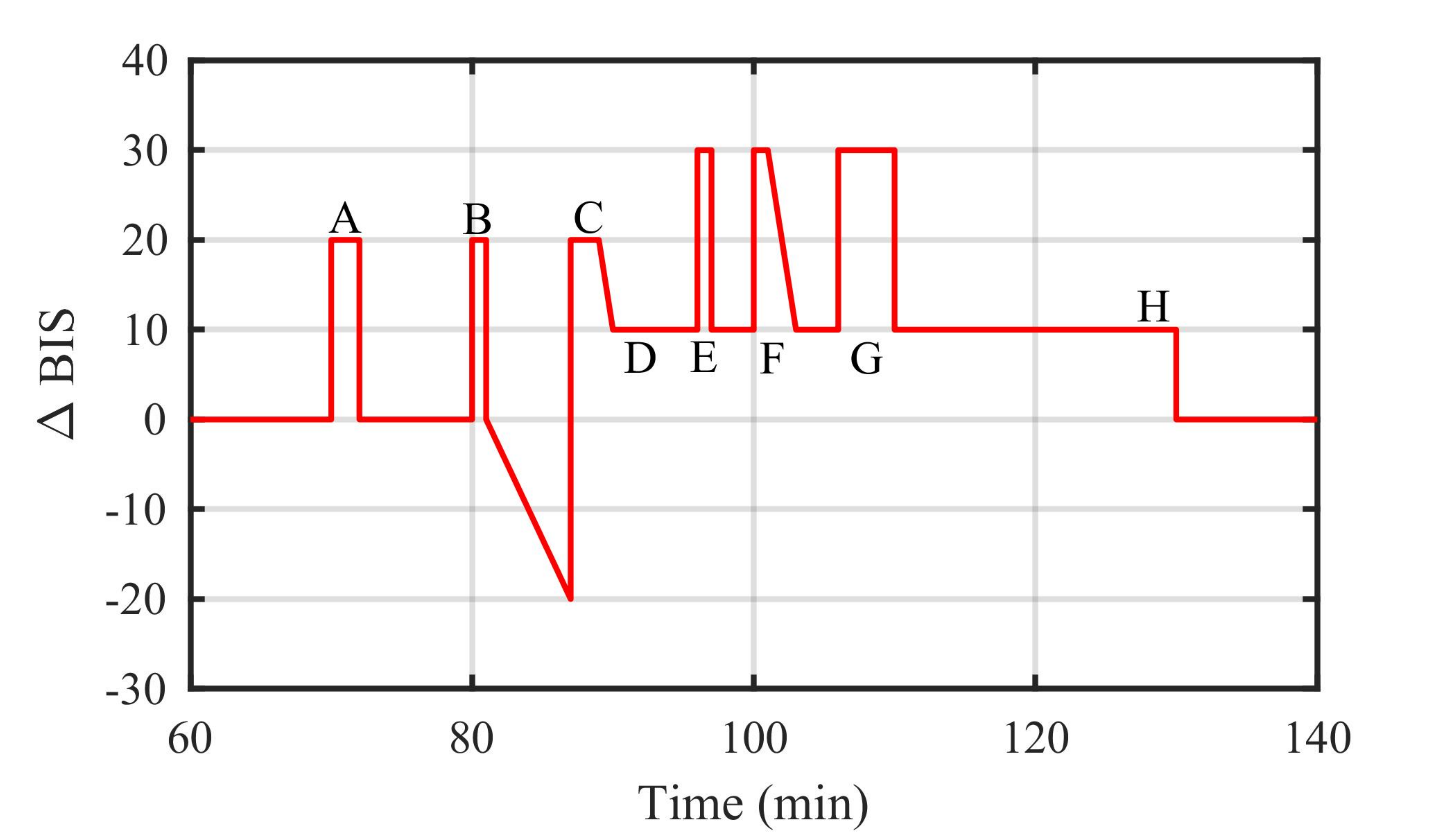}
\caption{A standard surgical stimulations profile.}
\label{fig12}
\end{figure}

\begin{figure}
\centering
\includegraphics[width=80mm]{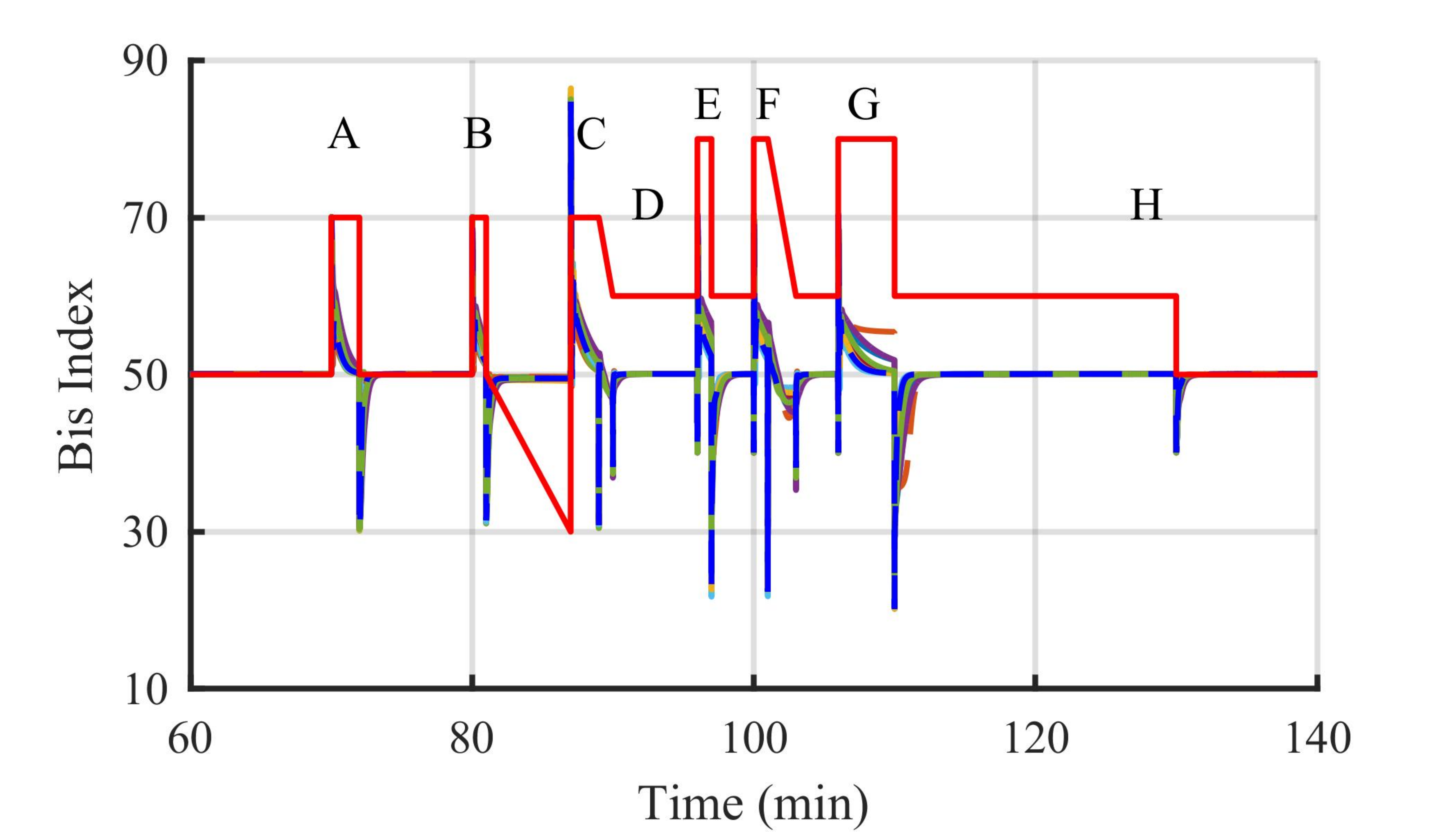}
\caption{BIS output for all patients in the maintenance phase (patient 9: orange dash line, nominal patient: blue dash line).}
\label{fig13}
\end{figure}

\begin{figure}
\centering
\includegraphics[width=80mm]{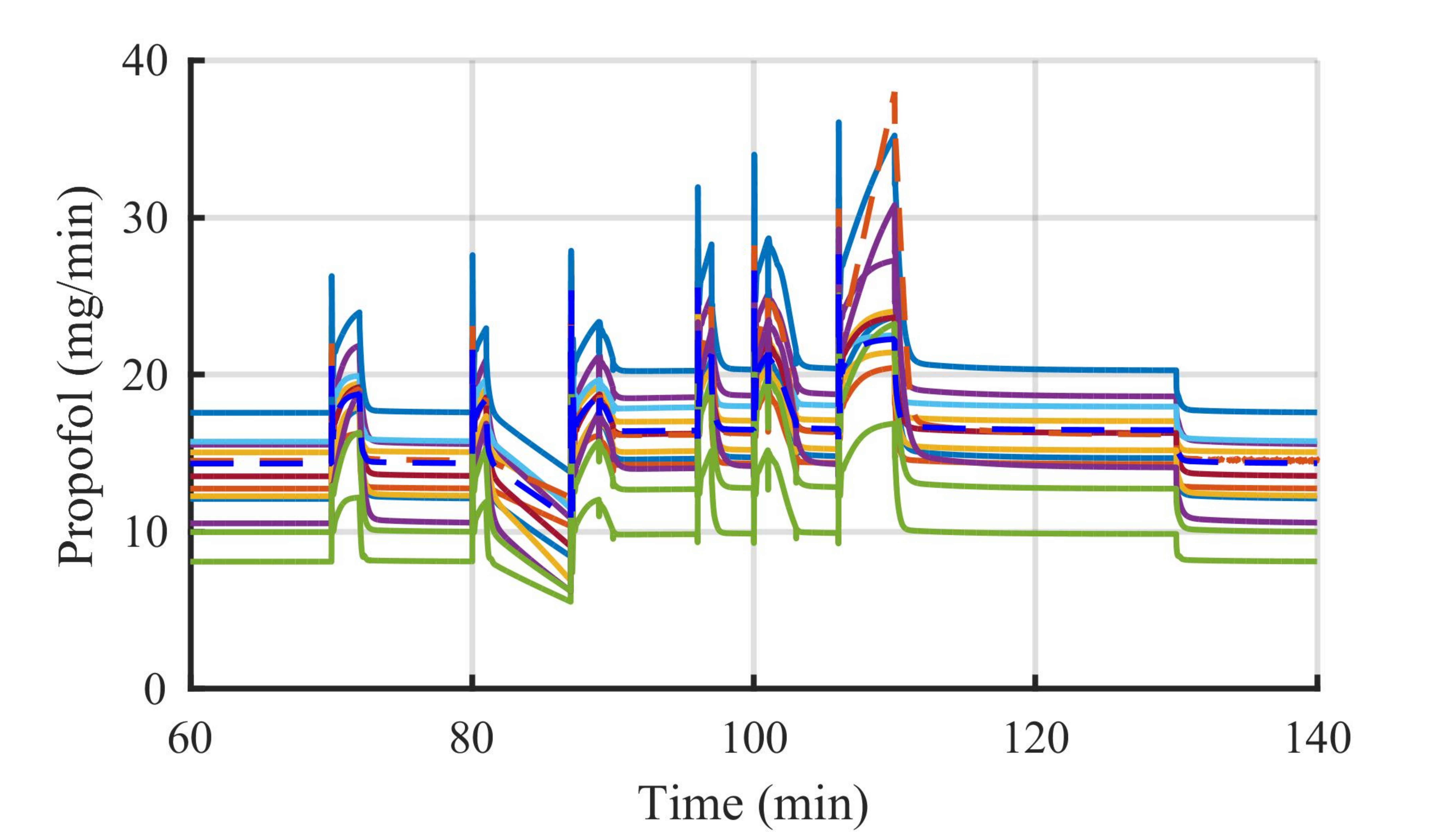}
\caption{Drug Infusion for all patients in the maintenance phase.}
\label{fig14}
\end{figure}

\begin{figure}
\centering
\includegraphics[width=80mm]{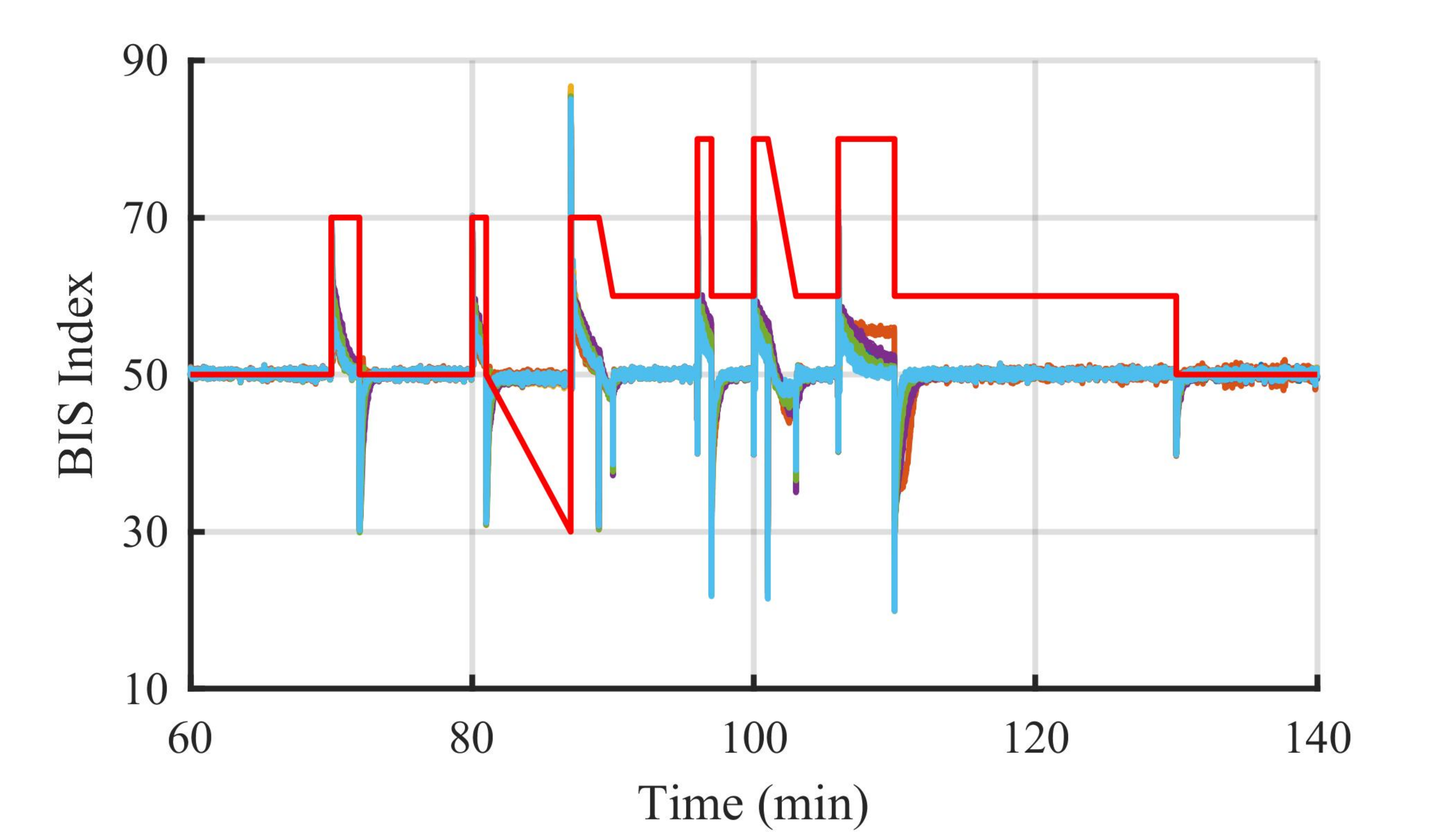}
\caption{Noise effect on BIS output for all patients in the maintenance phase.}
\label{fig15}
\end{figure}

\begin{figure}
\centering
\includegraphics[width=80mm]{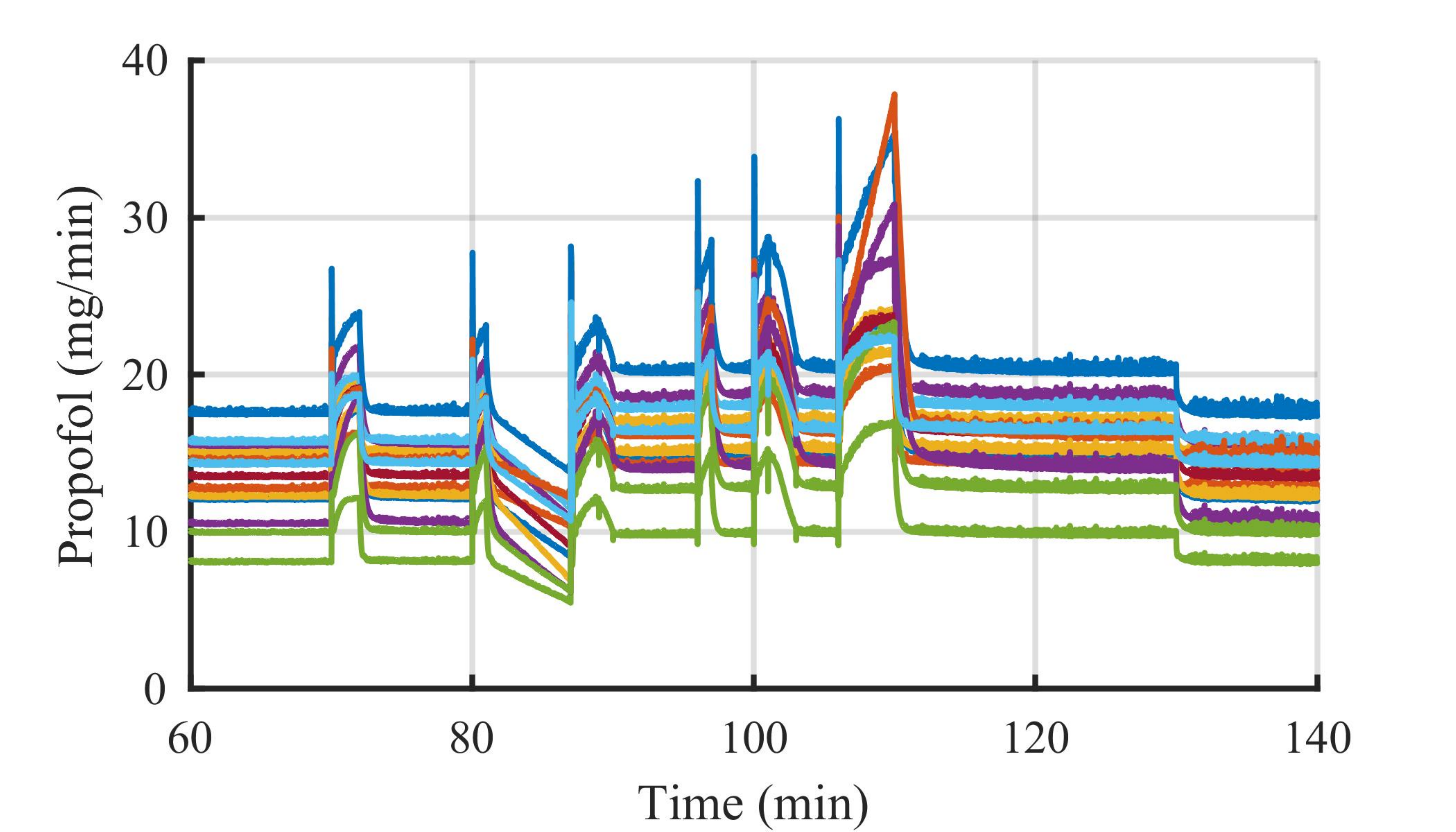}
\caption{Drug Infusion for all patients in the maintenance phase in the presence of noise.}
\label{fig16}
\end{figure}

As noted, the second mission of the controller is keeping the BIS value constant in the maintenance phase. In this section, this feature of the controller is evaluated by applying standard surgical stimulations and noises to the system output. Here, as shown in \textcolor{blue}{Fig. \ref{fig12}}, a standard surgical stimulation profile is employed to model the disturbances in the simulation\textcolor{blue}{\cite{struys2004performance}}. This profile contains some important steps in general surgeries, which effect on the BIS value. In \textcolor{blue}{Fig. \ref{fig12}}, $A$ shows the arousal due to laryngoscopy/intubation; $B$ represents surgical incision followed by a period of no surgical stimulation (e.g., waiting for a pathology result); $C$ represents an abrupt stimulus after a period of low-level stimulation; $D$ expresses the onset of a continuous normal surgical stimulation; $E$, $F$, and $G$ simulate short-lasting, larger stimulation within the surgical period; and $H$ denotes the withdrawal of stimulation during the closing period in\textcolor{blue}{\cite{struys2004performance}}. \textcolor{blue}{Fig. \ref{fig13}} and \textcolor{blue}{Fig. \ref{fig14}} shows that these surgical stimulations are applied to the model of all patients including the nominal patient without the noise while \textcolor{blue}{Fig. \ref{fig15}} and \textcolor{blue}{Fig. \ref{fig16}} show the results in the presence of the noise. The results prove that the controller is able to reject the surgical disturbances and conceal the noises by sub-optimal control action.  

To show the strength of the proposed controller, the results are compared with some MPC and PID controllers which have been recently used in closed-loop control of anesthesia. The proposed controller in the current paper can provide a fast response with no overshoot and undershoot in BIS response while the MPC, robust and non-overshooting controllers recently presented by\textcolor{blue}{\cite{nascu2017modeling, sadati2018multi, padmanabhan2019nonovershooting}} have longer settling time and cause overshoot in BIS response. Furthermore, here, the performance indexes given in\textcolor{blue}{\cite{struys2004performance}} are used, to have an analytical comparison and deeper evaluation of the presented controller. These indexes are integrated absolute error ($IAE$), performance error ($PE$), median performance error ($MDPE$), median absolute performance error ($MDAPE$), total variation ($TV$), $Q$ (quantity of administered drug), and WOBBLE (an index of response variations over time) and can be calculated as

\begin{equation}
    \label{eq24}
    IAE=\int_{0}^{\infty}\left | \overline{BIS}-BIS(t)) \right |dt
\end{equation}

\begin{equation}
    \label{eq25}
    TV=\sum_{k=0}^{\infty}\left | u_{k}-u_{k-1} \right |
\end{equation}

\begin{equation}
    \label{eq26}
    PE_{ij}=\frac{BIS_{j}-\overline{BIS}}{\overline{BIS}}\times100 \;\; \;\;\;\;\;\;\;j=1,...,N_j
\end{equation}

\begin{equation}
    \label{eq27}
    MDPE_i=Median\{PE_{ij}\;\;\;\;j=1,...,N_j\}
\end{equation}

\begin{equation}
    \label{eq28}
    MDAPE_i=Median\{|PE_{ij}|\;\;\;\;j=1,...,N_j\}
\end{equation}

\begin{equation}
    \label{eq29}
    WOBBLE_i=Median\{|PE_{ij}-MDAPE_i|\;\;j=1,...,N_j\}
\end{equation}
where $\overline{BIS}$ is the desired BIS value,  $BIS(t)$ is the measured BIS, $u_k$ is the current drug infusion rate, $u_{(k-1)}$ is the previous drug infusion rate, $i$ denotes the patient number, $j$ is the sample number, and $N_j$ defines the number of $PE$ values of each patient. \textcolor{blue}{{Table} \ref{table2}} shows the mean values and standard deviations of these indexes reported in\textcolor{blue}{\cite{padula2017optimized}} and\textcolor{blue}{\cite{merigo2017event}} and obtained results in this paper. As can be seen, the $AMCNFC$ can provide lower $MDPE$, which means it prevents overdosing. Since there is no overshoot, the value of $MDAPE$ is same as $MDPE$ for the proposed controller and has a lower value than two other controllers. This shows that $AMCNFC$ is a tighter controller and can reduce the periods of excessive anesthesia or reduce risk of awareness\textcolor{blue}{\cite{merigo2017event}}. Lower $MDAPE$ results in lower $WOBBLE$, which is more preferable from the clinical point of view. Also, as can be seen, the $AMCNFC$ has lower $IAE$ and $Q$ values due to the absence of overshoot in BIS response and ability of providing a sub-optimal control action. As can be seen in \textcolor{blue}{Table \ref{table2}}, the total variation of the proposed controller is acceptable, while the event based PID controller presented in\textcolor{blue}{\cite{merigo2017event}} has slightly smaller $TV$ due to establishing event based concept in the PID controller.

\section{Conclusion}\label{conclusion}

In this paper, the applicability of an adaptive neuro-fuzzy strategy in the closed-loop control of anesthesia has been evaluated numerically. The designed controller is evaluated in both induction and maintenance phases and proved its ability to provide an acceptable drug infusion rate for reaching the desired DOH. The results show that there is no overshoot and undershoot around the desired DOH and it is reached by a sub-optimal drug infusion in the desired time range. To evaluate the controller’s ability in dealing with inter and intra patient variability challenge, the proposed controller is applied to the PK/PD model of the 12 patients. The results indicate that the controller is adaptive and robust against this challenge and the measurement devices’ noises due to its independence from model parameters. The simulation results also show that the controller is able to reject the conventional surgical disturbances in the maintenance phase and can improve closed-control of anesthesia.      

\bibliographystyle{IEEEtran}
\bibliography{main.bib}

\end{document}